\documentclass[12pt,a4paper]{article}
\usepackage{amssymb}
\usepackage{afterpage}
\usepackage{citesort}
\renewcommand{\phi}{\varphi}

\newcommand{\be}{\begin{equation}}
\newcommand{\ee}{\end{equation}}
\newcommand{\ba}{\begin{eqnarray}}
\newcommand{\ea}{\end{eqnarray}}
\newcommand{\fr}{\frac}
\newcommand{\by}{\begin{array}}
\newcommand{\ey}{\end{array}}

\newcommand{\fullsquare}{{$\scriptstyle\square$}}
\newcommand{\fulltriangle}{{$\scriptstyle\triangle$}}

\begin{document}

\section*{\LARGE
Pair correlation functions in one-dim\-en\-sional
cor\-related-hop\-ping models}

\begin{flushleft}
{\large\bf M.~Quaisser, A.~Schadschneider, J.~Zittartz\\}
{\small\it
Institut f\"ur Theoretische Physik, Universit\"at zu K\"oln,
Z\"ulpicher~Str.~77, \mbox{D--50937 K\"oln}, Germany;\\
 E-mail:  mq@thp.uni-koeln.de, as@thp.uni-koeln.de, zitt@thp.uni-koeln.de\\}
\end{flushleft}\bigskip\bigskip

 {\noindent PACS. 75.10L, 74.20}\bigskip\bigskip

{\small\noindent {\bf Abstract.} We investigate ground-state
  properties of two correlated-hopping electron models, the Hirsch and
  the Bariev model.  Both models are of recent interest in the context
  of hole superconductivity. Applying the Lanczos technique to small
  clusters, we numerically determine the binding energy, the spin gaps,
   correlation functions, and other properties for various values of
  the bond-charge interaction parameter. Our results for small systems
  indicate that pairing is favoured in a certain parameter range.
  However, in contrast to the
  Bariev model, superconducting
  correlations are suppressed in the Hirsch model,
  for a bond-charge repulsion larger than a critical
  value.}\bigskip\bigskip

Since the discovery of high-$T_c$-superconductivity strong efforts
have been made to explain the pairing mechanism with the aid of
microscopic models for strongly correlated fermions, the Hubbard and
$t$-$J$ model being the most popular among them. In one dimension,
exact results can be derived for the Hubbard model
 and for the supersymmetric
$t$-$J$ model by applying the Bethe
ansatz \cite{korepin}. The main interest is the question whether and how
collective
behaviour can lead to a compensation of the Coulomb repulsion between
the electrons. As a possible explanation, models with
correlated-hopping interactions are a subject of current research
\cite{Hirsch1,Bariev,bksz1,bksz2,japa,fortunelli,buzatu,argentina,aligia,eta,etaschads,asbounds,quaisser2,beduerftig,karnaukhov,australier,BKZ}.

In 1989, Hirsch~\cite{Hirsch1} proposed a model for the description of
oxide superconductors by considering the holes in a nearly filled band
as the  charge carriers. The  Hamiltonian
contains among other contributions a correlated hopping interaction, i.~e.
a bond-charge repulsion.  A modified version of Hirsch's model, the
Bariev model, has been solved by Bethe ansatz~\cite{Bariev}. Although
it only contains one half of the bond-charge interaction terms of the
original model, it is expected to maintain its basic qualities because
it also takes into account the modification of the hopping amplitude
by the presence of a particle with opposite spin.
Considered as an electron model, the bond-charge repulsion leads to
the formation of Cooper pairs of holes, a process which is favoured at
small hole density \cite{bksz1}.

In this paper we compare exact diagonalization results for the two
models obtained by using the Lanczos
technique~\cite{gubernatis}. Although the systems
treated are small (up to 16 sites) and despite the finite size effects
which render the interpretation of the results difficult, comparison
with some exact results for the Bariev model (in the thermodynamic
limit), obtained from the Bethe ansatz solution, shows good
agreement.

In the following we shall consider a one-dimensional system consisting
of $N=N_\uparrow+N_\downarrow \leq 2L$ itinerant electrons on a closed chain
of $L$~sites with periodic (PBC) or antiperiodic (ABC) boundary
conditions. Electron hopping is possible between nearest
neighbour sites, but  the hopping matrix
element is modified by the presence of electrons with the opposite
spin direction on the sites involved in the hopping process.

The Hamiltonian of the Hirsch model reads \be\label{ham_Hirsch} {\cal
  H_H}(\Delta t)=-\sum_{j=1}^L \sum_{\sigma=\pm 1}\left( c_{j,\sigma}^+
  c_{j+1,\sigma} +c_{j+1,\sigma}^+ c_{j,\sigma}\right)\left(
  1-\fr{\Delta t}{2}\,\left(n_{j,-\sigma} +n_{j+1,-\sigma}\right)\right), \ee

and the Bariev model is described by the Hamiltonian
\be\label{ham_Bariev} {\cal H_B}(\Delta t)=-\sum_{j=1}^L \sum_{\sigma=\pm
  1}\left( c_{j,\sigma}^+ c_{j+1,\sigma} +c_{j+1,\sigma}^+
  c_{j,\sigma}\right)\left(
  1-\Delta t\,n_{j+\fr{1+\sigma}{2},-\sigma}\right), \ee where
$c_{j,\sigma}^+$ and $c_{j,\sigma}$, respectively, denote the creation
and annihilation operators of an up-electron ($\sigma=1$) or
down-electron ($\sigma=-1$) at site~$j$. $n_{j,\sigma}$ (which equals
0 or 1) is the number of particles with spin~$\sigma$ at site~$j$.
As can easily be seen, the interaction parameter $\Delta t$ is positive in
the repulsive model~\cite{bksz1} because it leads to a decrease of the
hopping probability (for $\Delta t<2$), if an electron with opposite spin
direction is present. In this sense
there is an additional correlation between the up-spin and down-spin
hopping processes.

\begin{figure}\begin{center}
\setlength{\unitlength}{10mm}
\begin{picture}(14.7,4)
\includegraphics{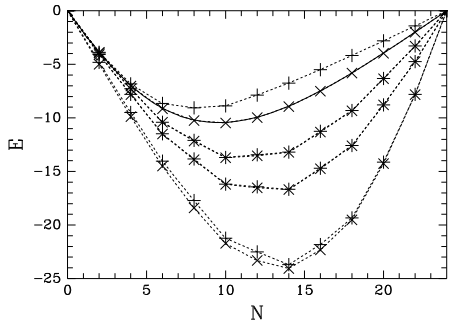}
\end{picture}
\caption{Ground-state energy, $L=12$ sites, of the Hirsch ($+$)
  and the Bariev model ($\times$) with PBC. $\Delta t=-1, -0.2,
  0.2, 1$ from  bottom to top. The analytic result for $\Delta t =1$
  is indicated as full line.}
\label{fg0001}
\end{center}
\end{figure}
%
%
%
We start our exact diagonalization study with the discussion of the
ground-state energy.  The energy difference between the Hirsch and the
Bariev model is very small for small $\Delta t$. As shown in
fig.~\ref{fg0001}, the energy curves, as functions of the particle
number $N$, coincide for $\Delta t=-0.2$ and $\Delta t=0.2$. They are also in
good
agreement for $\Delta t=-1$, but the curves spread apart for $\Delta t=1$,
where the
difference between the bond-charge interaction terms in the two models
becomes relevant. The figure also indicates that the energy difference
is small at low electron and low hole density for all values of the
interaction parameter $\Delta t$. For comparison, we included the exact
result \cite{quaisser2} for the Bariev model at $\Delta t=1$ as a full line,
which is $\frac{E}{L}=-\fr{2}{\pi}(1+\fr{n}{2})\sin(
\fr{n\pi}{1+\fr{n}{2}})$. The exact diagonalization results for $L=12$
 all lie on this curve.\\

%
\begin{figure}\begin{center}
    \setlength{\unitlength}{10mm}
\begin{picture}(14.7,4)
  \includegraphics{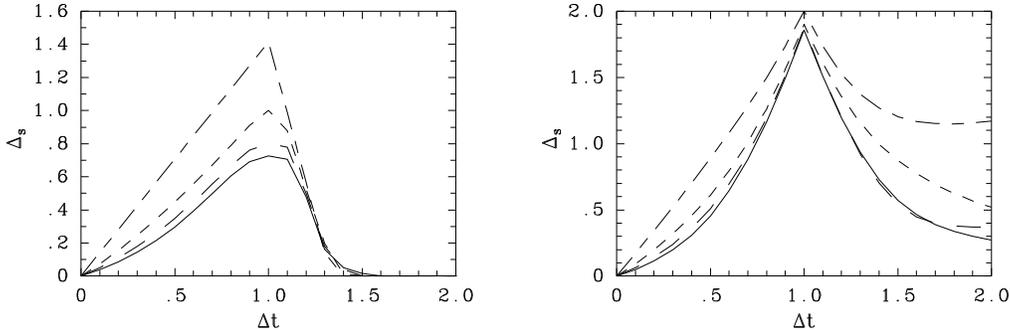}
\end{picture}
\caption{Spin gap of the Hirsch model (left) and Bariev model (right) for
$n=3/2$ and $L=4, 8, 12, 16$
  from top to bottom.}
\label{fg0014}
\end{center}
\end{figure}
%
%
%
%
%
In the following, we discuss properties of the Hirsch and Bariev model
for small hole filling. In the case of quarter filling of holes (i.~e.
electron filling $n=3/2$), we can compare results for systems with
$L=4$, 8, 12, and 16 sites.\\ The spin excitation energy is defined as
$\Delta_s=E(S^z=1)-E(S^z=0)$, where $E(S^z)$ is the ground-state
energy in the subspace with fixed $S^z$; it corresponds to the lower critical
magnetic
field $h_c$ which is necessary to create a non-zero magnetization in
the system. In the absence of a spin excitation gap, $h_c$ vanishes,
as is the case in the repulsive Hubbard model
and the supersymmetric $t$-$J$ model.  On
the
contrary, the attractive Hubbard model  and the
Bariev model have a gap in the spin
excitations (see \cite{bksz2,Myletter,quaisser2} and references therein).\\ In
fig.~\ref{fg0014}  we show the results for $\Delta_s$
for both models. We choose
PBC for $L=16$ and 8 and ABC for $L=$12 and 4, because the results
facilitate the extrapolation into the thermodynamic limit if the Fermi
wavenumber $k_F=n\pi/2$ of the free infinite system is one of the
possible $k$ numbers of the finite system~\cite{spronken}. We verified
that there is already good agreement between the exact
diagonalization results for $L=12$ and the analytic solution
$\Delta_s=\fr{2}{\pi}\sin\left(\fr{n\pi}{1+\fr{n}{2}}\right)-
\fr{4n}{2+n}\cos\left(\fr{n\pi}{1+\fr{n}{2}}\right)$ \cite{quaisser2}
for the Bariev model at $\Delta t=1$.  The results strongly suggest that the
spin gap is maximal at $\Delta t=1$ in both models and that the spin gap is
vanishing in the Hirsch model for $\Delta t\gtrsim 1.5$ in contrast to the
Bariev
model where it remains non-zero.  The vanishing of $\Delta_s$ at $\Delta t=2$
in
the Hirsch model is evident for any system size and boundary condition
and a large range of densities. It is due to the high degeneracy at
that parameter value where the Hirsch model can be treated
analytically and the ground-states are of the $\eta$-pairing
type~\cite{eta,etaschads}. Moreover,  results for other  hole
densities indicate that the vanishing of the spin gap in the Hirsch
model is not restricted to the band filling of $n=3/2$.

Whenever there is a gap in the spin excitation spectrum, spin
correlations decay exponentially so that other correlations, such as pair
correlations are more likely to dominate. We calculate the singlet
pair correlation functions for on-site and next-neighbour pairs and
the triplet pair correlation function, defined as
\begin{eqnarray*}
  G_0(r)&=&\langle P^+_0(r)P_0(0)\rangle,\\ G_1(r)&=&\langle
  P^+_s(r)P_s(0)\rangle,\\ G_2(r)&=&\langle P^+_t(r)P_t(0)\rangle,
\end{eqnarray*}
with the pairing operators
\begin{eqnarray*}
  P^+_0(r)&=&c^+_{r\uparrow}c^+_{r\downarrow},\\
  P^+_s(r)&=&\frac{1}{\sqrt{2}}(c^+_{r\uparrow}c^+_{r+1,\downarrow}
  -c^+_{r\downarrow}c^+_{r+1,\uparrow}),\\
  P^+_t(r)&=&\frac{1}{\sqrt{2}}(c^+_{r\uparrow}c^+_{r+1,\downarrow}
  +c^+_{r\downarrow}c^+_{r+1,\uparrow}).
\end{eqnarray*}

\begin{figure}\begin{center}
    \setlength{\unitlength}{10mm}
\begin{picture}(14.7,4)
  \includegraphics{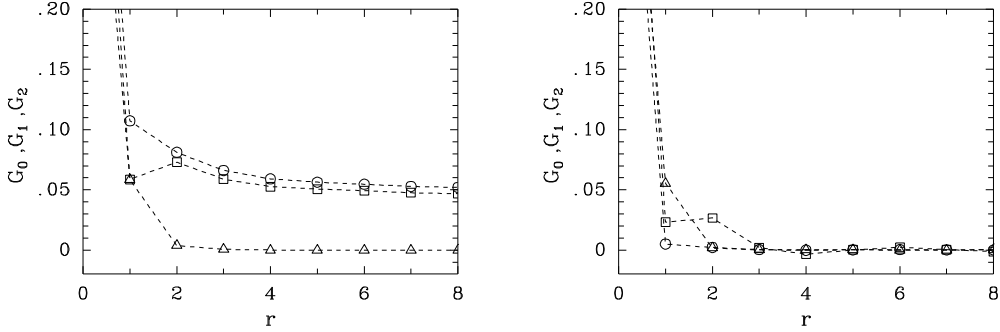}
\end{picture}
\caption{Spatial singlet pair correlations $G_0$ ($\circ$), $G_1$
  (\fullsquare), and triplet pair correlations $G_2$ (\fulltriangle)
  of the Hirsch model  at $\Delta t=1$ (left) and $\Delta t=1.7$ (right) for
$L=16$, $n=3/2$.}
\label{fg0015}
\end{center}
\end{figure}
\begin{figure}\begin{center}
    \setlength{\unitlength}{10mm}
\begin{picture}(14.7,4)
  \includegraphics{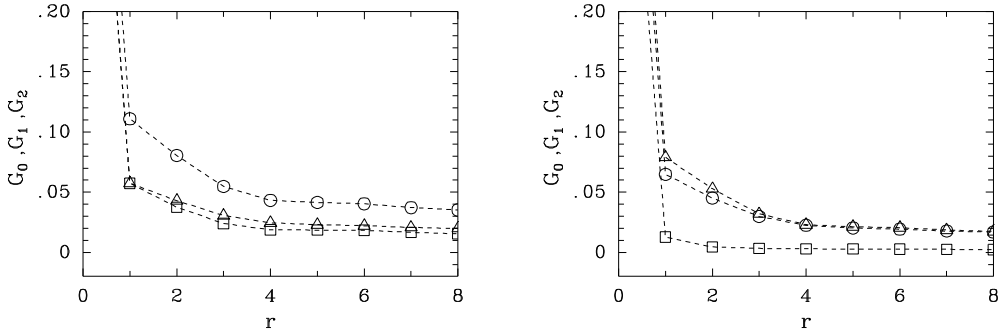}
\end{picture}
\caption{Same as in fig.~\ref{fg0015} for the Bariev model.}
\label{fg0016}
\end{center}
\end{figure}
When the number of doubly occupied sites is reduced in the
ground-state of the interacting system---as compared to free
fermions---$G_1(r)$ is expected to be larger than $G_0(r)$, although
both functions should be equivalent with respect to their decay
 in the infinite system. We compare results for both models
for $\Delta t=1$ and $\Delta t = 1.7$.  Fig.~\ref{fg0015} shows that at
$\Delta t=1$ both singlet correlation functions of the Hirsch model decay
slowly to a non-zero value (which nevertheless can be zero in the
infinite system, when finite size effects are absent). The triplet
correlations decay very rapidly to zero. On the other hand, all pair
correlations decay fast to zero at $\Delta t=1.7$ where there is no spin
excitation gap. This is different for the Bariev model where the decay
is slower (fig.~\ref{fg0016}). We mention that the  spin
operator $\hat{S}^2$ does not commute with the Hamiltonian ${\cal
  H_B}$ (\ref{ham_Bariev}) so that $S$ is no good quantum number in
the Bariev model, in contrast to the Hirsch model.
\afterpage{\clearpage}

\begin{figure}\begin{center}
    \setlength{\unitlength}{10mm}
\begin{picture}(14.7,4)
  \includegraphics{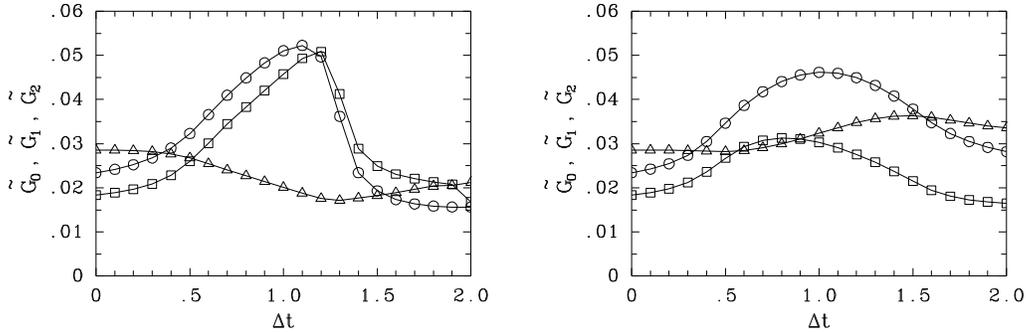}
\end{picture}
\caption{Fourier transform $\tilde{G}(k)$ of the pair correlation
  functions of the Hirsch model (left) and Bariev model (right), at $k=0$.
$\tilde{G}_0$ ($\circ$),
  $\tilde{G}_1$ (\fullsquare), $\tilde{G}_2$ (\fulltriangle).}
\label{fg0017}
\end{center}
\end{figure}

To gain some insight into the dependence of the pair correlation
functions on the bond-charge interaction parameter $\Delta t$, we also show
their Fourier transform $\tilde{G}_j(k)=\frac{1}{L}\sum_r
G_j(r)\cos(kr)$ ($j=0,1,2$) for $k=0$ which is simply the sum of the
correlations over all spatial distances. Although it is not possible
to discuss the decay  by means of this function only, it
helps to determine at which values of the interaction pair
correlations are favoured.  Fig.~\ref{fg0017}  indicates that singlet
pairing is strong at $\Delta t\approx 1$ in both models, i.~e.\ where the
spin gap is large. This tendency is diminished drastically in the
Hirsch model at parameters $\Delta t\gtrsim 1.1$. On the other hand, these
functions are decreasing more slowly in the Bariev model. The curves
of $\tilde{G}_0$ and $\tilde{G}_1$ have a similar shape and only differ
by magnitude. Triplet correlations are small in both models and vary
little as functions of $\Delta t$.

\begin{figure}\begin{center}
    \setlength{\unitlength}{10mm}
\begin{picture}(14.7,4)
  \includegraphics{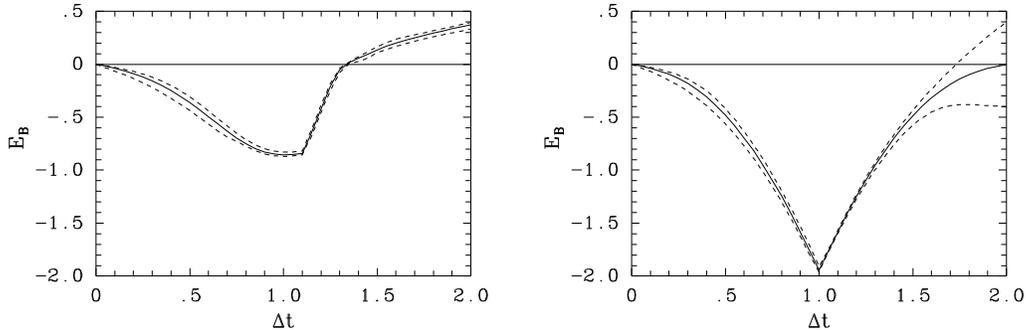}
\end{picture}
\caption{Binding energy of the Hirsch model (left) and Bariev model (right),
$n=20/12$, averaged over
  TBC (full line). The broken lines estimate the finite size
  deviations.}
\label{fg0018}
\end{center}
\end{figure}
To complete the comparison of the two models, we calculate the binding
energy $E_B(N)=E(N+2)+E(N)-2E(N+1)$. To suppress the finite size effects, we
choose twisted boundary conditions (TBC), which means that the wavefunction
at site 0 and site $L$ is related by $\Psi(L)=e^{i\phi}\Psi(0)$, and
take the average of $E_B$ over various values of the twisting angle
$\phi$.  To estimate the finite size effects, we also show the
maximum/minimum $E_B$ obtained in this way (fig.~\ref{fg0018}). The
binding energy is negative in the Hirsch model for $\Delta t<1.4$, but
becomes positive for larger values of $\Delta t$. In the Bariev model,
the averaged $E_B$ is zero at $\Delta t=0$ and $\Delta t=2$ and
negative for all other $\Delta t$, with minimum at $\Delta t=1$. For
$\Delta t>1.7$, the finite size effects in the Bariev model are too
large to decide if binding occurs or not, but we have strong evidence
that binding is suppressed in the Hirsch model for $\Delta
t>1.5$.\smallskip

We conclude that our calculation of various ground-state properties of
finite clusters at small hole doping indicates that the Bariev model
and the Hirsch model show similar behaviour as long as the bond-charge
repulsion $\Delta t$ is small. Pair correlations are strongest in both
models for $\Delta t\approx 1$.  For $\Delta t\gtrsim 1.5$, there is a
change in the ground-state properties of the Hirsch model, and
superconducting correlations are strongly suppressed for large $\Delta
t$, which is in accordance with mean field results~\cite{buzatu}. In
the Bariev model, there is no such evident change.  We finally propose
a simple argument which might help to understand why the parameter of
$\Delta t\approx 1.5$ can be a phase boundary in the Hirsch model: The
analytic diagonalization of the local interaction (acting only on two
neighbouring sites) of (\ref{ham_Hirsch}) shows that (for $1<\Delta
t<2$) $|\Psi_1\rangle=|\sigma 2\rangle + |2 \sigma\rangle$ and
$|\Psi_2\rangle=\left|\uparrow\downarrow\rangle\right.-\left|\downarrow\uparrow\rangle\right.
+ |20\rangle +|02\rangle$ (0, $\sigma$, 2 denoting the empty, singly,
and doubly occupied site) are local eigenvectors with eigenvalues
$E_1=1-\Delta t$ and $E_2=-2+\Delta t$~\cite{asbounds}. Since
$E_1=E_2$ for $\Delta t=3/2$, there is a level-crossing in the local
ground-state at that parameter, whereas this effect is absent in the
Bariev model.  This criterion might be of importance for the global
ground-state, too.\bigskip

{\noindent\bf Acknowledgments}\bigskip

This work was performed within the research program of the
Sonderforschungsbereich 341, K\"oln-Aachen-J\"ulich. We thank the HLRZ
at KFA J\"ulich for giving access to the supercomputing facilities and
  A.~A.~Aligia and K.~Hallberg for helpful discussions.

{\clearpage}

\suppressfloats

\end{document}